# BASIC, SIMPLE AND EXTENDABLE KINETIC MODEL OF PROTEIN SYNTHESIS

*Alexander N. Gorban*[1,2], *Annick Harel-Bellan*[3,4], *Nadya Morozova*[3,4], and *Andrei Zinovyev*[5,6,7,*]

[1] University of Leicester, Center for Mathematical Modeling, Leicester, UK

[2] Lobachevsky University, Nizhny Novgorod, Russia

[3] Institute for Integrative Biology of the Cell (I2BC), CEA, CNRS, Université Paris- Saclay, 91198, Gif-sur-Yvette cedex, France.

[4] Institut des Hautes Etudes Scientiques, Bures-sur-Yvette, France

[5] Institut Curie, 26 rue d'Ulm, F-75248 Paris France

[6] INSERM U900, France

[7] Mines PariTech, Fontainbleau, France

[*]Corresponding author

E-mail: andrei.zinovyev@curie.fr

## Abstract

Protein synthesis is one of the most fundamental biological processes, which consumes a significant amount of cellular resources. Despite existence of multiple mathematical models of translation, varying in the level of mechanistical details, surprisingly, there is no basic and simple chemical kinetic model of this process, derived directly from the detailed kinetic model. One of the reasons for this is that the translation process is characterized by indefinite number of states, thanks to existence of polysomes. We bypass this difficulty by applying a trick consisting in lumping multiple states of translated mRNA into few dynamical variables and by introducing a variable describing the pool of translating ribosomes. The simplest model can be solved analytically under some assumptions. The basic and simple model can be extended, if necessary, to take into account various phenomena such as the interaction between translating ribosomes, limited amount of ribosomal units or regulation of translation by microRNA. The model can be used as a building block (translation module) for more complex models of cellular processes. We demonstrate the utility of the model in two examples. First, we determine the critical parameters of the single protein synthesis for the case when the ribosomal units are abundant. Second, we demonstrate intrinsic bi-stability in the dynamics of the ribosomal protein turnover and predict that a minimal number of ribosomes should pre-exists in a living cell to sustain its protein synthesis machinery, even in the absence of proliferation.

# Introduction

Production of proteins is one of most fundamental cellular processes, taking up to 75% of cellular resources in terms of chemical energy, in simple microbes [1]. The transcription-translation process description, mentioning only the most basic "elementary" processes, consists in:

1) production of mRNA molecules (including splicing),
2) initiation of these molecules by circularization with help of initiation factors,
3) initiation of translation, recruiting the small ribosomal subunit,
4) assembly of full ribosomes,
5) elongation, i.e. movement of ribosomes along mRNA with production of protein,
6) termination of translation,
7) degradation of mRNA molecules,
8) degradation of proteins

Despite quite "linear" description of this process, a difficulty in the kinetic modeling of it arises when one tries to take into account the phenomenon of polysome [2,3], when several ribosomes are synthesizing peptides on a single mRNA at the same time. This leads to multiplicity of possible states of mRNA with various ribosome numbers and potentially different dynamics, interaction between ribosomes and other complex phenomena. This difficulty was evident already in the first published mathematical models of protein synthesis [4–6].

The process of protein synthesis and translation is a subject of mathematical modeling since long time ago starting from detailed kinetic models [4,5,7], taking into account stochastic aspects of translation [8] and using computer simulations for the case of large polysomes [9]. A number of chemical kinetics-based models of protein synthesis have been developed and analyzed in the last four decades [10–12]. Beyond chemical kinetics, various modeling formalisms such as Totally Asymmetric Simple Exclusion Process (TASEP) [13–15], Probabilistic Boolean Networks (PBN), [16], Petri Nets and max-plus algebra [17] have been applied to model the detailed kinetics of protein synthesis or some of its stages [18].

In the context of specific questions and applications of mathematical modeling to translation, dealing with detailed kinetic description of translation might be not optimal, and simplified models of translation can become more suitable. Thus, few attempts have been made in order to simplify the detailed kinetic description of protein synthesis. For example, two simple kinetic models of translation were introduced before in [19] and analyzed in detail in [20]. However, they were introduced without strict derivation from the detailed translation kinetics and did not allow taking into account neither degradation of mRNA nor existence of polysomes. Ad hoc simplified models of protein synthesis have been exploited for addressing specific contexts of translation regulation [21–23]. Using simplified models allows more direct determination of the most important control parameters of protein translation regulation.

We share the point of view that "useful models are simple and extendable" [24]. Following this paradigm, one needs to create the simplest kinetic model of protein synthesis and suggest a way to complexify it if needed to address a particular observation. Despite very long history of the mathematical modeling of protein synthesis, to our knowledge, no *basic and simple* kinetic description of the process, directly and formally derived from its detailed representation, was suggested until so far. This is the gap we close in this study.

In the following we start with a 1) **detailed mechanistic description** of the translation process with explicit representation of every state of translated mRNA, followed by 2) deriving the **simplest and basic kinetic model** of coupled transcription, translation and degradation, and 3) **extending the model** in order to take into account various effects. In this paper, the extensions will describe the saturation of mRNA initiation rate, effects of ribosome interactions, regulation of translation by microRNA.

The basic model is constructed by 1) correct lumping of the detailed model states and by 2) separating the descriptions of ribosomal turnover and the translation initiation through introducing a variable representing the pool of translating ribosomes. The simplest model remains linear under assumption of that the local concentrations of ribosomal subunits or initiation factors remain constant or changes relatively slowly. To avoid non-physiological properties (such as a possibility of infinite number of ribosomes per mRNA), we modify the model by introducing delays in the initiation of ribosome and the effects of ribosome interactions. In this form, the model becomes more realistic but non-linear in some extensions.

## Results

*Detailed kinetic model of translation*

Let us introduce the following notations:

$L$ – length of mRNA (in nucleotides);

$l_m$ – length occupied on mRNA by fully assembled ribosome (in nucleotides);

$k_t$ – rate constant of production of mRNA molecules;

$k_d$ – rate constant of degradation of mRNA molecules;

$k_r$ – speed of movement of translating ribosome along mRNA (nucleotise/sec);

$IF$ – various initiation factors;

S40 – small ribosome component;

S60 – large ribosome component.

Further we will use squared brackets to denote the concentrations of the corresponding molecular species: for example, [S40] will denote the local concentration of S40 ribosomal subunits. For the amounts of the components we keep the same notations as for the components themselves. Thus, $R$ is amount of amount of ribosomes and the total amount of mRNA molecules is *MT*.

The simplest assumption about the production and destruction of mRNA is that the degradation process does not depend on the state of mRNA. Under this assumption the total pool of mRNAs is produced at rate $k_t$ and destroyed with rate constant $k_d$, i.e. its dynamics is simple and autonomous:

$$\frac{dMT}{dt} = k_t - k_d \times [MT] \,.$$

It is worth to notice that the production rate $k_t$ is an extensive quantity (it scales with the total volume of the system) whereas all rate constants are intensive ones.

The total pool of mRNA molecules can be separated in sub-pools of mRNA molecules in different states:

$R_0$ – mRNA molecules in non-initiated state (not ready for translation)

$\underline{R}_0$ – mRNA molecules in initiated state (ready for translation, with 40S subunit sitting at the mRNA)

$R_1$ – mRNA molecules with one single ribosome assembled and moving along the mRNA

$\underline{R}_1$ – mRNA molecules with one ribosome assembled and initiated for new incoming ribosome

$R_2$ – mRNA molecules with two ribosomes assembled and moving along the mRNA

$\underline{R}_2$ – mRNA molecules with two ribosomes assembled and initiated for new incoming ribosome

….

$R_{nmax}$ – mRNA molecules with *nmax* ribosomes assembled and moving along the mRNA

$\underline{R}_{nmax}$ – mRNA molecules with *nmax* ribosomes assembled and initiated for new incoming ribosome

The sum of all sub-pools of mRNA should be equal to *MT*:

$$MT = \sum_{i=0}^{nmax}(R_i + \underline{R}_i)$$

The number *nmax* is defined as the maximum number of ribosomes able to sit on mRNA: it may be roughly evaluated as

$nmax = L / l_m$.

Schematically, the process of translation can be represented as in Figure 1.

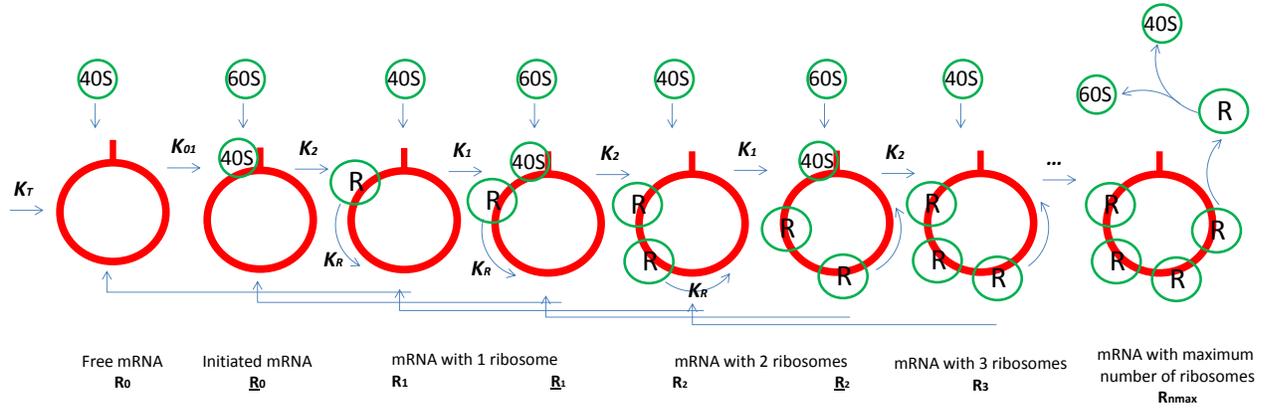

Figure 1. Schematic process of detailed translation representation. It requires 2 x (*nmax*+1) mRNA states.

The time of passage of one ribosome along mRNA may be evaluated as

$t_p = L/k_r$ ,

hence, the reaction rate constant of protein production and subsequent release of ribosomes from mRNA (shown in Figure 1 by backward arrows) may be evaluated as:

$k_3 = k_r/L$ .

The transformation of states is described by the following chemical equations:

$R_i \to \underline{R_i}$ (with rate constant $k_1$), $i = 0\ldots nmax$

$\underline{R_i} \to R_{i+1}$ (with rate constant $k_2$), $i = 0\ldots nmax-1$

$R_i \to R_{i-1}$ (with rate constant $k_3$), $i = 1\ldots nmax$

$\underline{R_i} \to \underline{R_{i-1}}$ (with rate constant $k_3$), $i = 1\ldots nmax$

$R_i \to R_{i-1}$ (with rate constant $k_{rd}$), $i = 1\ldots nmax$

$\underline{R_i} \to \underline{R_{i-1}}$ (with rate constant $k_{rd}$), $i = 1\ldots nmax$

## *Basic model of protein synthesis, constructed by lumping the states of the detailed model*

To avoid using 2×(*nmax*+1) states (which potentially can be large) to represent translation, we lump the description of the detailed process in the following way. We denote

*M* – amount of mRNA with translation initiation site not occupied by assembling ribosome,

*F* – amount of mRNA with translation initiation site occupied by assembling ribosome,

*R* – amount of ribosomes sitting on mRNA synthesizing proteins,

*P* – amount of proteins.

In terms of $\underline{R_i}$ and $R_i$ variables, $M$ and $F$ represent the lumped values:

$$M = \sum_{i=0}^{nmax} R_i, \quad F = \sum_{i=0}^{nmax} \underline{R_i} \text{ and } MT = M + F.$$

There are two lumped reactions and two reactions representing the turnover of ribosomes (as a result of translation termination and protein synthesis or spontaneous ribosome drop-off from mRNA without protein production):

$M \rightarrow F$ with reaction rate constant $k_1$,

$F \rightarrow M + R$ with reaction rate constant $k_2$,

$R \rightarrow$ null with reaction rate constant $k_3$.

$R \rightarrow$ null with reaction rate constant $k_{rd}+k_d$ (ribosome drop-off and degradation without protein production).

The reaction network describing transcription, translation and mRNA degradation is represented in Figure 2. We will denote this model as $\mathcal{M}0$.

The corresponding list of equations is

$$\begin{cases} \dot{M} = k_t - k_d M - k_1 M + k_2 F \\ \dot{F} = k_1 M - k_d F - k_2 F \\ \dot{R} = k_2 F - k_3 R - k_{rd} R - k_d R \\ \dot{P} = k_3 R - k_p P \end{cases} \quad (1)$$

which has the following solution for zero initial condition $M(0) = F(0) = R(0) = P(0) = 0$

$$M(t) = \frac{k_t}{k_d} \frac{k_2(k_1 + k_2 + k_d)(1 - e^{-k_d t}) + k_1 k_d (1 - e^{-(k_1+k_2+k_d)t})}{(k_1 + k_2 + k_d)(k_1 + k_2)},$$

$$F(t) = \frac{k_t}{k_d} \frac{k_1(k_1 + k_2 + k_d)(1 - e^{-k_d t}) - k_1 k_d (1 - e^{-(k_1+k_2+k_d)t})}{(k_1 + k_2 + k_d)(k_1 + k_2)},$$

$$R(t) = \frac{k_t}{k_d} \frac{k_1 k_2}{(k_3 + k_{rd} + k_d)(k_1 + k_2 + k_d)}$$
$$\times \left[ 1 - \frac{(k_3 + k_{rd} + k_d)(k_1 + k_2 + k_d)}{(k_3 + k_{rd})(k_1 + k_2)} e^{-k_d t} + \frac{(k_3 + k_{rd} + k_d)k_d}{(k_3 + k_{rd} - k_1 - k_2)(k_1 + k_2)} e^{-(k_1+k_2+k_d)t} + \frac{(k_1 + k_2 + k_d)k_d}{(k_3 + k_{rd} - k_1 - k_2)(k_3 + k_{rd})} e^{-(k_3+k_{rd}+k_d)t} \right],$$

$$P(t) = \frac{k_t}{k_d} \frac{k_1 k_2 k_3}{k_p(k_1 + k_2 + k_d)(k_3 + k_{rd} + k_d)} \left[ 1 - \alpha_d e^{-k_d t} + \alpha_{12} e^{-(k_1+k_2+k_d)t} + \alpha_3 e^{-(k_3+k_{rd}+k_d)t} - \alpha_p e^{-k_p t} \right], \quad (2)$$

$$\alpha_d = \frac{(k_3 + k_{rd} + k_d)(k_1 + k_2 + k_d)}{(k_3 + k_{rd})(k_1 + k_2)} \frac{k_p}{k_p - k_d}, \quad \alpha_{12} = \frac{(k_3 + k_{rd} + k_d)k_d}{(k_3 + k_{rd} - k_1 - k_2)(k_1 + k_2)} \frac{k_p}{k_p - k_d - k_1 - k_2},$$

$$\alpha_3 = \frac{(k_1+k_2+k_d)k_d}{(k_3+k_{rd}-k_1-k_2)(k_3+k_{rd})} \frac{k_p}{k_p-k_d-k_3-k_{rd}}, \quad \alpha_P = 1-\alpha_d+\alpha_3+\alpha_{12}$$

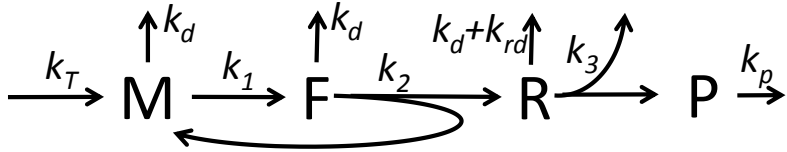

Figure 2. Basic and the simplest model ℳ0 of protein synthesis.

The simplest model ℳ0 can be made more complex if some particular aspects of translation are needed to be represented in more details. Below we build several such modifications. In the model ℳ0' we explicitly model the first round of mRNA initiation which can be longer than the consequent rounds of 40S recruitment and production of translating ribosomes in the pool. In the model ℳ1 we explicitly model the step of binding of 40S and 60S subunits to mRNA. In the model ℳ1' we also explicitly add the binding of the initiation factors. In the model ℳ0'reg we introduce the effect of irreversible binding of a regulatory molecule to mRNA which can be, for example, a microRNA.

### ℳ0' *model*: *Distinguishing the initial initiation stage in the basic model*

An assumption implicitly made in the simplest model ℳ0 is that the process of the first translation initiation (on a just transcribed mRNA) takes the same amount of time as consequent translation initiations on the mRNA already having translating ribosomes. In reality, the time needed to process transcribed mRNA into the form ready for translation can take significant time, including such steps as splicing, circularization, etc.

In order to model this additional initial delay, specific states of mRNA such as $R_0$ (free mRNA) and $\underline{R}_0$ (initiated mRNA) can be separately represented in the model. Let us denote the amount of mRNA in these states as $M_0 = R_0$ and $F_0 = \underline{R}_0$. The corresponding reaction network is shown in Figure 3. This model is able to represent specific states of just produced, non-initiated mRNA. This model contains two additional parameters: $k_{01}$ and $k_{02}$, which are rate constants of the first round of mRNA initiation and firing the first assembled ribosome into the pool. Evidently, these constants cannot be smaller than $k_1$ and $k_2$, because they include some additional events: $k_1$ corresponds to recruiting 40S while $k_{01}$ corresponds to initiating the new-born mRNA and recruiting 40S on it. Thus, typically $k_{01} << k_1$.

If $k_{02} << k_2$ then this can also represent translation with membrane-bound ribosomes or SRP cycle (Singh, 1996), when there is a transient translation arrest in the initiated monosome state (the very beginning of the translation).

Separating $M_0$ and $F_0$ states also allows estimating the average number of ribosomes RB sitting on an *initiated* mRNA (the pool represented by $M$ and $F$ states in $M0'$).

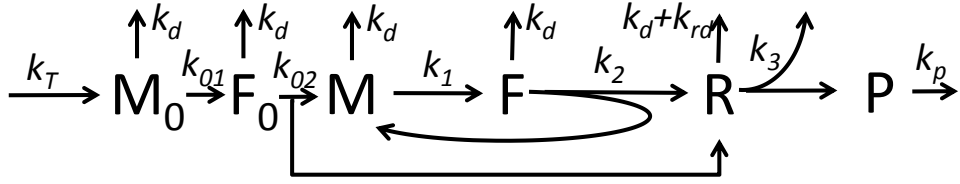

Figure 3. Reaction network representing translation process with explicit representation of the fraction of initiated $F_0$ and non-initiated, free (more exactly, "early born") mRNA $M_0$ (model $M0'$).

The corresponding system of equations is

$$\begin{cases} \dfrac{dM_0}{dt} = k_t - (k_d + k_{01})M_0 \\ \dfrac{dF_0}{dt} = k_{01}M_0 - (k_d + k_{02})F_0 \\ \dfrac{dM}{dt} = k_{02}F_0 + k_2 F - (k_d + k_1)M \\ \dfrac{dF}{dt} = k_1 M - (k_d + k_2)F \\ \dfrac{dR}{dt} = k_{02}F_0 + k_2 F - (k_d + k_{rd} + k_3)R \\ \dfrac{dP}{dt} = k_3 R - k_p P \end{cases}$$

(3)

which has the following steady-state solution:

$$M_0 = \frac{k_t}{k_{01} + k_d}, \quad F_0 = \frac{k_t k_{01}}{(k_{01} + k_d)(k_{02} + k_d)},$$

$$M = \frac{k_t}{k_d} \frac{k_{01} k_{02}(k_2 + k_d)}{(k_{01} + k_d)(k_{02} + k_d)(k_1 + k_2 + k_d)}, \quad F = \frac{k_t}{k_d} \frac{k_{01} k_{02} k_1}{(k_{01} + k_d)(k_{02} + k_d)(k_1 + k_2 + k_d)},$$

$$R = \frac{k_t}{k_d} \frac{k_{01} k_{02}(k_1 + k_d)(k_2 + k_d)}{(k_{01} + k_d)(k_{02} + k_d)(k_1 + k_2 + k_d)(k_3 + k_d + k_{rd})},$$

$$P = \frac{k_3}{k_p} \frac{k_t}{k_d} \frac{k_{01} k_{02}(k_1 + k_d)(k_2 + k_d)}{(k_{01} + k_d)(k_{02} + k_d)(k_1 + k_2 + k_d)(k_3 + k_d + k_{rd})},$$

$$MT = M_0 + F_0 + M + F = \frac{k_t}{k_d},$$

$$RB = \frac{R}{M+F} = \frac{(k_1+k_d)(k_2+k_d)}{(k_1+k_2+k_d)(k_3+k_d+k_{rd})}. \quad (4)$$

The relaxation times are

$$rt_{M_0} = \frac{1}{k_{01}+k_d}, \; rt_{F_0} = \frac{1}{\min(k_{01}+k_d, k_{02}+k_d)}, \quad (5)$$

$$rt_M = rt_F = \frac{1}{k_d}, \; rt_R = \frac{1}{k_d}, \; rt_P = \frac{1}{\min(k_d, k_p)}.$$

The *M-F* subsystem has the kinetic matrix with eigenvalues $-k_d$ and $-k_d-k_1-k_2$.

## Models $\mathcal{M}1$ and $\mathcal{M}1'$: *Explicit representation of 40S, 60S and initiation factors binding*

One of the undesired features of the simplest translation models $\mathcal{M}0$ and $\mathcal{M}0'$ is a possibility of unrealistic increase of the number of translating ribosomes in the pool. The kinetic rate constants $k_1$ and $k_2$ implicitly include the concentrations (not amounts) of 40S and 60S subunits correspondingly. Increasing these concentrations might lead to the unlimited growth of the steady-state amount of ribosomes (2).

Therefore, in order to create a more detailed and realistic representation of reaction $M \to F$, one can include the intermediate step of reversible binding of mRNA to the small ribosomal subunit $M + 40S \to M{:}40S$ and the scanning step during which 40S bound to mRNA search for the start codon: $M{:}40S \to F$. Here *F* represents a state of mRNA with 40S positioned at the start codon and ready to recruit 60S. The time needed for finding the start codon ($\sim 1/k_a$) is a complex function of the local concentrations of certain initiation factors and, possibly, length and the secondary structure of 5'UTR.

Similar to $\mathcal{M}0'$, we can decouple the two initial states of mRNA in $\mathcal{M}1$ and produce the model $\mathcal{M}1'$, in which binding of initiation factors (*IF*1 and *IF*2) can be represented explicitly (Figure 5). In this model we distinguish two types of initiation factors: *IF*1 initiate mRNA by binding to the cap structure, poly-A tail, etc.; *IF*2 initiate assembly of ribosomes and can be RNA helicases or other helper molecules [25]. *IF*1 factors are released only when the initiated states of mRNA (all besides $M_0$) are degraded. *IF*2 are released in the end of each ribosome assembly.

It is important to make a notice on the usage and recycling of 40S, 60S and IFs. All these molecules make a pool of resources (together with ATP and GTP, aminoacids, tRNAs, etc.) shared between many protein syntheses in the whole cell. The equations (1), (3) are written down for the amounts of the corresponding proteins, while 40S, 60S and *IF*s are consumed with the rates proportional to their local *concentrations* (Figure 5). 40S, 60S and *IF*s molecules are returned to the pool of cellular resources in four ways: 1) in each act of mRNA degradation

(except for the just transcribed $M_0$ state of mRNA) with rate constant $k_d$, 2) release of ribosomes from mRNA with rate constants $k_3$ and $k_{rd}$, 3) in backward reactions of 40S detachment from mRNA with rate constants $k_1^-, k_{01}^-$ (not shown explicitly in Figure 5), 4) in releasing a new translating ribosome with the rate constants $k_2, k_{02}$.

We assume that each individual protein synthesis does not significantly change the pool of cellular resources and, therefore, the local concentrations of 40S, 60S and *IFs* remain constant. With such quite a realistic assumption, the models remain linear and analytically tractable. However, this might not be completely satisfactory approximation for the in vitro cell-free systems for studying translation, when the amounts of 40S or 60S or *IFs* are made comparable to the amounts of the translated mRNA. In this case recycling of ribosomal subunits and initiation factors might be a limiting (and fast) process, thus it should be represented explicitly, taking into account the effective volume occupied by 40S or 60S or *IFs* in the system (because the kinetic rates of resources release give the amount of the released translation factors while their consumption rates are proportional to their concentrations).

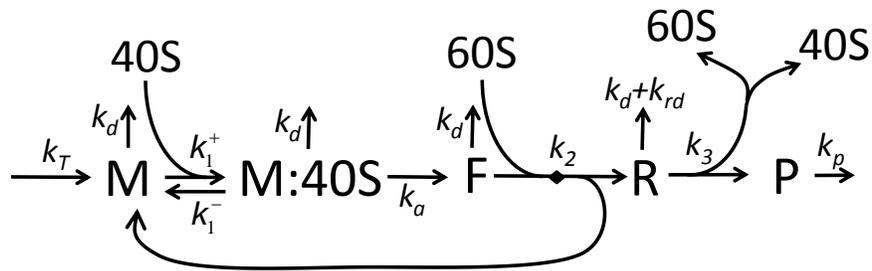

Figure 4. Reaction network representing translation process with explicit presentation of 40S and 60S binding (model ℳ1).

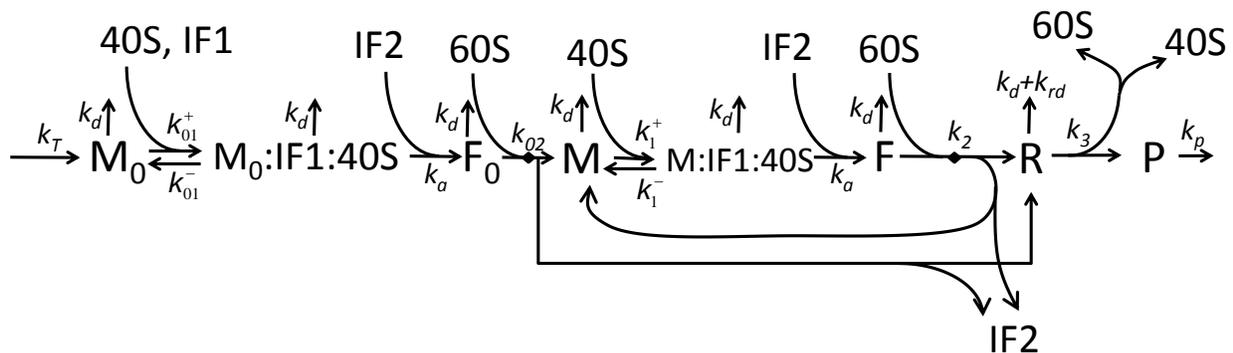

Figure 5. Reaction network representing translation process with explicit presentation of 40S, 60S and initiation factors (IF) binding (model ℳ1').

The steady state solution of the model $\mathcal{M}1$, assuming that the ribosomal units 40S and 60S are available in excess is:

$$M^{SS} = \frac{k_t}{k_d} \frac{k_2[60S] + k_d}{\frac{k_1^+ k_2}{k_1^- + k_a + k_d}[40S][60S] + k_d + k_2[60S] + \frac{k_1^+(k_a + k_d)}{k_1^- + k_a + k_d}[40S]}$$

$$F^{SS} = \frac{k_t k_1^+ k_a}{k_d} \frac{[40S]}{k_1^+ k_2[40S][60S] + k_d(k_1^- + k_a + k_d) + k_2(k_1^- + k_a + k_d)[60S] + k_1^+(k_a + k_d)[40S]}$$

$$(M:40S)^{SS} = \frac{k_t k_1^+}{k_d} \frac{[40S](k_2[60S] + k_d)}{k_1^+ k_2[40S][60S] + k_d(k_1^- + k_a + k_d) + k_2(k_1^- + k_a + k_d)[60S] + k_1^+(k_a + k_d)[40S]}$$

$$R^{SS} = \frac{k_t}{k_d} \frac{k_a}{k_3 + k_d} \frac{[40S][60S]}{[40S][60S] + \frac{k_d(k_1^- + k_a + k_d)}{k_1^+ k_2} + \frac{k_1^- + k_a + k_d}{k_1^+}[60S] + \frac{k_a + k_d}{k_2}[40S]} \quad (6)$$

$$P^{SS} = \frac{k_t}{k_d} \frac{k_a}{k_3 + k_d} \frac{k_3}{k_p} \frac{[40S][60S]}{[40S][60S] + \frac{k_d(k_1^- + k_a + k_d)}{k_1^+ k_2} + \frac{k_1^- + k_a + k_d}{k_1^+}[60S] + \frac{k_a + k_d}{k_2}[40S]}$$

*Model $\mathcal{M}10'$reg: extending the basic model of translation with microRNA-based regulation*

Let us assume that the translation process is regulated by a molecule which can irreversibly bind to mRNA and, as a result, can change one or several kinetic rates of translation. Typical example of such a molecule is microRNA [26], so we will call it like this further. MicroRNAs are short stretches of RNA, able to regulate translation of the majority of human proteins, playing an important role in normal physiological processes and diseases such as cancer [27,28].

For our purposes (representing microRNA-based regulation), it is important to distinguish states $M_0$ and $F_0$ to be able to represent the initiation of mRNA and the effect of microRNA on the initiation process. MicroRNA can act on $k_{01}$ step ($M_0 \rightarrow F_0$), thus inhibiting the early initiation process, or on $k_1$ step ($M \rightarrow F$), thus, inhibiting step of 40S binding on already initiated mRNA, or on $k_2$ step ($F \rightarrow M+R$), thus inhibiting ribosome assembly process [20,26,29].

To take into account the action of microRNA on translation, the model of translation shown in Figure 3 is supplied with mRNA states representing mRNA with a microRNA bound to it (states $M'_0$, $F'_0$, $M'$, $F'$, $R'$). The rate of microRNA binding is $k_b$ which determines irreversible conversion of the microRNA-free states (without prime) to microRNA-bound states (primed). The corresponding rate constants which might be different from normal translation process are marked with prime symbol as well. In addition, we introduce a special $B$ state which describes reversible capturing of mRNA in P-bodies, where they can be specifically degraded at a higher rate $k_{bd}$ than during the microRNA-free translation.

The $\mathcal{M}0'reg$ model was used in [26] to produce the kinetic signatures of nine different mechanisms of microRNA action or their combinations. It was shown that each of the nine possible mechanisms has its own characteristic kinetic signature, which gives to experimentalists a tool to discriminate between them in their particular experimental system. The provided characteristic kinetic signature of an individual mechanism represents a characteristic plot with the predicted dynamics of 3 measurable biochemical variables (mRNA concentration, the corresponding protein concentration, the average number of ribosomes at a translated mRNA) in the case when a microRNA act on a given mRNA via this exact mechanism only.

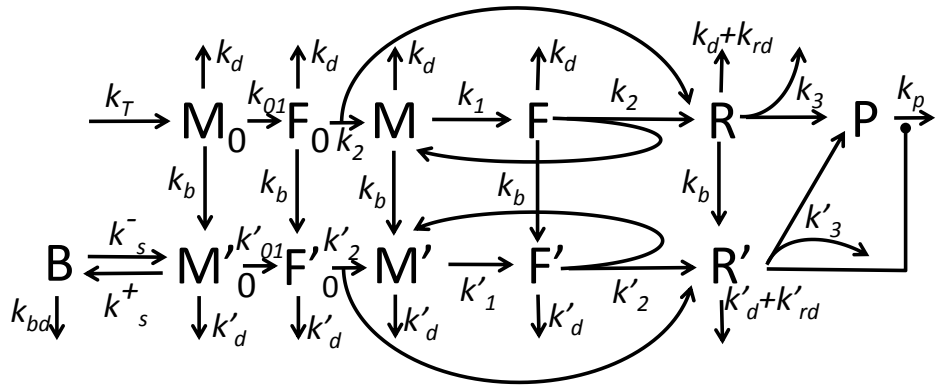

Figure 6. The model $\mathcal{M}0'reg$ of miRNA-based translation regulation [26].

*Other possible model extensions*

The basic lumped model can serve as a basis for other model extensions by explicit splitting of particular states from the lumped states and other modifications. Let us list several possible scenarios:

1) More explicit representation of translation termination or elongation, description of ribosome stalling phenomenon.

2) More detailed representation of the mRNA initiation process. For example, formation of the M0:IF1:40S complex in Figure 5 should proceed in several elementary steps, with particular role and order of binding of scaffold initiation proteins and other initiation factors, with subsequent recruitment of 40*S*.

3) Description of phenomena connected with uneven distribution of ribosomes along mRNA, such as described in recent literature on explicit studies of ribosome positioning on mRNAs (Ingolia et al, 2009).

4) Explicit modeling of the mRNA codon usage.

5) Mean-field models of the ribosomes' interaction: The simplest method to include the interaction of ribosomes in the lumped model is a dependence of the ribosome drop-off constant $k_{rd}$ on the average concentration θ of the ribosomes per initiated molecule of mRNA: $k_{rd}=k_{rd}(θ)$. For example, for the scheme presented in Figure 3 it may be $k_{rd}(θ)=a/(b–θ)$, where $θ=R/(M+F)$.

6) Mean-field models of how the property of the mRNA (such as its stability) might change depending on the state and also on the history of mRNA. For example, one can imagine a (very hypothetical) version of mRNA kinetics with "mRNA aging" such as each new round of translation makes mRNA more fragile and prone to destruction. Or, in opposite, mRNA can become more stable with ribosomes sitting on it.

7) Modeling distribution of model parameters, leading to existence of population of mRNAs with different speeds of different steps of translation.

8) Explicit modeling of competition of various protein syntheses processes for resources (ribosomal subunits and initiation factors). The most interesting is to include in this picture the production of the resources themselves (transcription, translation, degradation), which will introduce complex global regulatory feedbacks.

### *Example of application: determining factors limiting translation in the M1' model*

In order to illustrate what distinguishes two particular translation models described above, we performed a numerical experiment in which we varied the concentrations of ribosomal subunits and studied their effect on the average number of translating ribosomes per mRNA. We compared two models M0' and M1', without and with an intermediate state of mRNA bound to 40S ribosomal subunit but with 40S not yet positioned at the start codon. Our purpose is to demonstrate two points: 1) that the step of late initiation might be very sensitive parameter and lead to efficient regulation of translation (which is consistent with experimental findings [30]; 2) that without this step a simpler model M0' can lead to non-physiological unlimited growth of translating ribosomes per mRNA (Figure 7).

As one can see from Figure 7, the steady state value of the average number of translating ribosomes per mRNA is not limited in the model M0', if the concentrations of small and large ribosomal subunits are increased *simultaneously*. Increasing only concentration of 60*S* with fixed concentration of 40*S* is not sufficient: to increase the number of complexes one has to supply the system with both components.

By contrast, in the model M1', simultaneous increase in the concentrations of 40S and 60S makes *RB* insensitive of them (Figure 7, right plot). This can be easily understood from the model shown in Figures 4 and 5. If one assumes that the rates of mRNA degradation and synthesis are slower than the translation rate then it is easy to show that the steady-state value of the average number of translating ribosomes per mRNA is

$$RB = \frac{1}{k_3} \times \frac{k_a k_1^+ k_2 [IF1][IF2][40S][60S]}{k_2[60S](k_a[IF2]+k_1^-)+k_1^+[IF1][40S](k_2[60S]+k_a[IF2])}.$$

Hence, if both ribosomal subunits are in excess then

$$RB\Big|_{[60S],[40S]\to\infty} = \frac{k_a[IF2]}{k_3}. \qquad (7)$$

This is the limiting value of the average number of translating ribosomes per mRNA in the models $\mathfrak{M}1$ and $\mathfrak{M}1'$.

Formula (7) has an important biological consequence: in the excess of ribosomal subunits the most sensitive parameter of protein synthesis (which is determined by *RB*) is the availability (or, equivalently, efficiency) of the initiation factors facilitating fixation of 40S bound to mRNA at the start codon (collectively denoted as *IF*2 in Figure 6). Good candidates for this type of initiation factors are RNA helicases whose role is to disentangle the 5'UTR regions of mRNA [25,30]. The early initiation factors, collectively denoted as *IF*1 in Figure 6, can play less important role, if the ribosomal subunits are in excess (they do not enter into (7)).

If both initiation factors are in excess then there is saturation with respect to their values. For fixed concentrations of the ribosomal subunits we get

$$RB\Big|_{[IF1],[IF2]\to\infty} = \frac{k_2[40S][60S]}{k_3}.$$

Therefore, saturation with respect to ribosomal subunits and initiation factor concentrations is not symmetric: the limiting parameter value with respect to infinite increase of initiation factors depends on both 40*S* and 60*S* concentrations while the limiting value with respect to infinite increase of ribosomal subunits depends only on the concentration of *IF*2.

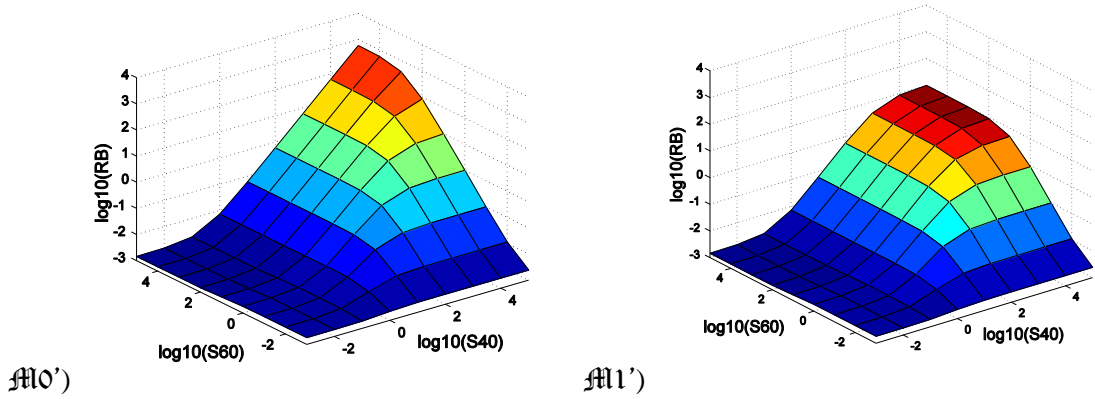

Figure 7. Number of translating ribosomes per mRNA (*RB*) in 𝔐0' and 𝔐1' models of translation as a function of concentrations of small (*S*40) and large (*S*60) ribosomal subunits for fixed concentrations of the initiation factors. The set of parameters used in this simulation is provided in the Methods and Materials section.

## *Example of application: modeling ribosomal protein synthesis*

Models developed in the previous sections can help understand general and global properties of protein synthesis in a living cell. Instead of focusing on a single protein synthesis, one can consider a global machinery of synthesis of all cellular proteins or some abundant groups of them.

Therefore, it is interesting to consider applying the basic model of protein synthesis to the model the synthesis of ribosomal proteins, because in this case there exists an intrinsic feedback mechanism regulating the amount of ribosomal proteins in a living cell.

At first, we exploited for this purpose model 𝔐1 (Figure 4) which suggests how in a quasi-steady state the number of translating ribosomes $R^{SS}$ depends on the concentrations of 40S and 60S components (see formula (6)). Assuming physiological translation (efficient binding of 40S to mRNA and not too strong mRNA degradation), we can put $k_a \gg k_1^-, k_d \ll k_a, k_d \ll k_3$ and simplify (6) to

$$R^{SS} = \frac{k_t}{k_d}\frac{k_a}{k_3}\frac{[40S][60S]}{[40S][60S]+\frac{k_d k_a}{k_1^+ k_2}+\frac{k_a}{k_1^+}[60S]+\frac{k_a}{k_2}[40S]} = \frac{k_a}{k_3}\frac{MR^S[40S][60S]}{[40S][60S]+nm+m[60S]+p[40S]},$$

(8)

where $MR^S = \frac{k_t}{k_d}$, $m = \frac{k_a}{k_1^+}, n = \frac{k_d}{k_2}, p = \frac{k_a}{k_2}$. The meaning of $MR^S$ parameter is the total number of mRNAs of a given protein or a protein type.

In order to study very general features of ribosomal protein synthesis, let us assume that [40S] and [60S] components are identical and denote them collectively as $S = [40S] = [60S]$ and use the rate equation (8):

$$\frac{dS}{dt} = k_a MR^S \frac{S^2}{S^2 + (p+m)S + mn} - k_s S \qquad (9)$$

The equation (9****) is characterized by a possibility of bistability, with two non-zero steady states (one of which is stable and another one is unstable):

$$S^{High} = \frac{1}{2}\left[ MR^S \frac{k_a}{k_s} - (p+m) + \sqrt{\left(MR^S \frac{k_a}{k_s} - (p+m)\right)^2 - 4mn} \right]$$

$$S^{Low} = \frac{1}{2}\left[ MR^S \frac{k_a}{k_s} - (p+m) - \sqrt{\left(MR^S \frac{k_a}{k_s} - (p+m)\right)^2 - 4mn} \right]$$

Three steady states of the protein synthesis system ($S=0$, $S=S^{High}$, $S=S^{Low}$) exist only when

$$\left( MR^S \frac{k_a}{k_s} - (p+m) \right)^2 \geq 4mn, \qquad (10)$$

which we can consider as a general condition of cell viability, because otherwise sustainable protein synthesis is not possible.

In the important asymptotic case $\left( MR^S \frac{k_a}{k_s} - (p+m) \right)^2 \gg 4mn$, one has

$$S^{High} \approx MR^S \frac{k_a}{k_s} - (p+m) = k_a\left( \frac{k_t}{k_d} \frac{1}{k_s} - \frac{1}{k_2} - \frac{1}{k_1^+} \right),$$

$$S^{Low} \approx \frac{mn}{MR^S \frac{k_a}{k_s} - (p+m)} = \frac{k_d}{\frac{k_t}{k_d} \frac{k_1^+ k_2}{k_s} - (k_1^+ + k_2)} . \qquad (11)$$

The meaning of $S^{Low}$ is the minimum number of ribosomes needed to maintain the translation (when it is possible, accordingly to the condition (10)). If the number of ribosomes drops below this threshold, then the translation will collapse to the zero steady state (Figure 8,A). Interestingly, in the asymptotic case (11), $S^{Low}$ decreases with increasing efficiency of translation initiation ($k_1^+$, $k_2$ coefficients) and does not depend on $k_a$. $S^{High}$, which meaning is the stable number of ribosomes in a cell, has the opposite behavior: it increases with more efficient translation and linearly scales with the scanning rate $k_a$. This behavior is illustrated in (Figure 8,B) for some realistic estimation of the protein synthesis parameters.

This analysis shows that for a biologically relevant set of parameters, the critical necessary minimal number of ribosomes in a cell seems not to exceed few thousands. It becomes larger in the case when the ribosomal protein translation is slow due to its rate limiting initiation. For large intervals of values of initiation rates, the critical amount can approach 1 ribosome, which means that theoretically, a cell can recover from drastic (close to complete) ribosome depletion, by synthesizing a new complete pool of them. As a strong speculation, one can suggest that evolutionary the parameters of translation are chosen such that a cell could robustly recover its protein synthesis machinery.

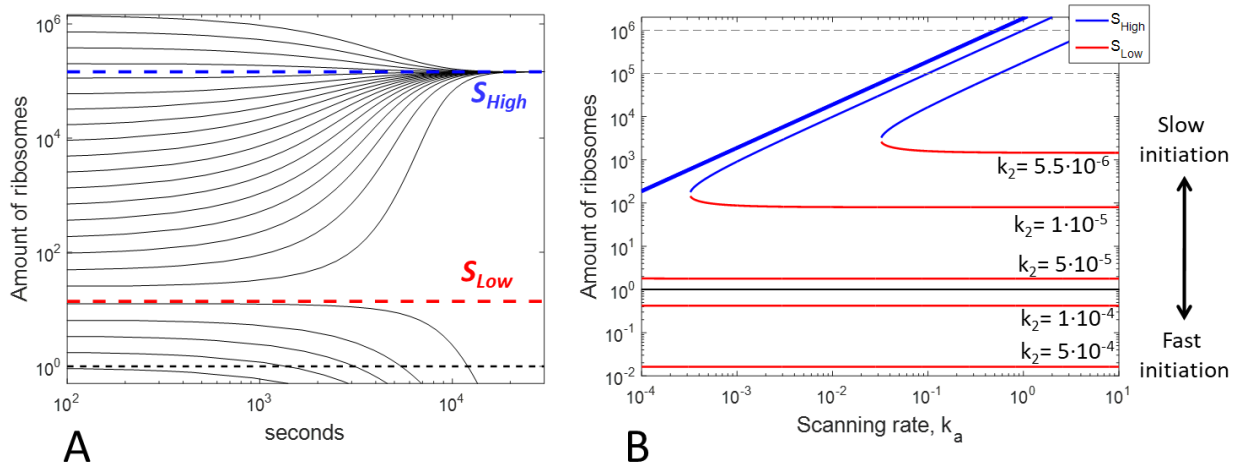

Figure 8. Simplest model of ribosomal production predicts existence of a critical number of ribosomes needed for cell survival. A) Time series showing the dynamics of the number of ribosomes $S$ from different initial amounts. Here parameters $k_t$, $k_d$, $k_s$ are estimated as described in the text, while $k_a = 0.1$, $k_2 = 2 \cdot 10^{-5}$, $k_1^+ = 0.1 k_2$. Here black dashed line denotes the amount in 1 ribosome, which is considered as the viability threshold. B) $S^{Low}$ and $S^{High}$ value dependence on the translation parameters. Dashed grey lines denote typical range of the number of ribosomes in a unicellular eukaryotic cell (yeast). $S^{Low} < 1$ denotes a regime in which a cell can theoretically recover its protein synthesis from close to complete depletion of ribosomes.

## Methods and materials

Numerical simulations were made using MATLAB. Executable model definitions are provided from http://github.com/sysbio-curie/ProteinTranslationModels .

Rough parameter estimation was made using numbers from [31] and http://book.bionumbers.org, using data for a simple eukaryotic cell, such as yeast. Typical mRNA half-life was assumed in 20-30 minutes, which is reflected in the value $k_d = 8 \cdot 10^{-4}$ sec$^{-1}$. The number of mRNA molecules for a particular protein was estimated on average in 1000, which leads in $k_t = 10^3 k_d$. Typical protein half-lie was assumed in 30 mins-1 hour which gives $k_s = 4 \cdot 10^{-4}$ sec$^{-1}$. We assumed the stable number of ribosomes in a cell in $2 \cdot 10^5$-$5 \cdot 10^5$ which constrains the value $S^{High}$ and the protein/mRNA ratio in $10^2$ by order of magnitude. For other parameter values ($k_1^+$, $k_2$, $k_a$) we did not fix the exact parameters but rather scan their ranges, assuming that the initiation of mRNA with 40S is faster than full ribosome assembly ($k_1^+ = 10\ k_2$) and that the 5' scanning step is relatively fast ($k_a \gg k_1^+$). We underline here that these parameter values' choice does not change the formulas derived in the manuscript and the conclusions about the model's dynamic behavior but rather used for illustrative purpose.

## Discussion

In this paper we derive a simple model of protein synthesis which is directly derived from a simplification of detailed kinetics of protein translation, describing the phenomenon of polysome. The simplification is achieved through lumping, one of the common approaches from reaction network asymptotology toolbox [32]. This derivation distinguishes the model from other simpified models of protein synthesis introduced *ad hoc* in various studies. Simple explanatory illustration of the nature of the suggested model is provided in Figure 9 ("opening/closing door"-type modeling).

The model is made extendable such that it makes it relatively easy to represent in more details some particular aspects of protein synthesis dynamics, if this is needed. We provide several model extensions, each of which can have specific applications. One of such extension deals with a feature of the basic model which might be non-desirable: namely, a possibility for unlimited number of ribosomes in a polysome. Explicit representation of a reaction step describing the scanning by the 40S ribosomal subunit for the start codon along the 5' end of mRNA limits the number of ribosomes in polysome and allows determining the most sensitive parameters of translation initiation under various assumptions (e.g., for the case of unlimited access of 40S and 60S subunits).

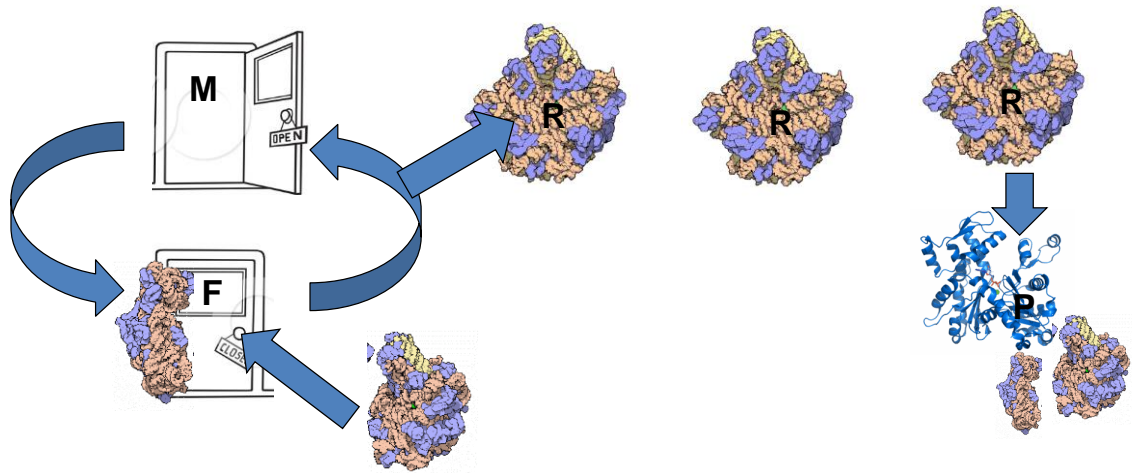

Figure 9. A mechanistic interpretation of the simplest model of translation $\mathcal{M}0$. A mRNA is a place of translation which can exist in "open" (*M*) and "closed" (*F*) states. When the place is "open" it can accept a small ribosomal subunit, after which the place is "closed" until the large ribosomal subunit is recruited and the assembled ribosome is released into the pool of translating ribosomes.

Another example of application that we provided is modeling the synthesis of ribosomal protein pool in a cell. We demonstrate that in this cellular system there exists an intrinsic bi-stability, with three steady-states for the amount of ribosomes in a cell, two stable ones, 0 and $S^{High}$, and one unstable, $S^{Low}$. We derive formulas for these values connecting them to the basic parameters of protein synthesis. The biological meaning of $S^{Low}$ is the minimum number of ribosomes required for a cell in order to sustain its protein synthesis. If the number of ribosomes drops below this number, then the whole machinery of protein synthesis is predicted to collapse, not being able to maintain the synthesis of ribosomal protein pool. Interestingly, the usual estimates for the number of ribosomes in a living cell is made for the case of actively proliferating cells using simple arguments for the necessity of protein pool replenishment in dividing cells [31]. By contrast, the estimates for $S^{High}$ (stable number of ribosomes) and $S^{Low}$ (minimum number of ribosome state from which the protein synthesis is able to recover) are valid even for quiescent cells. It is known that the house-keeping proteins have relatively long half-lives: therefore, sustaining cellular life can less crucially depend on the *de novo* protein synthesis and availability of the ribosomal proteins.

For quiescent cells, the experiment with depleting the ribosomal protein pool might be feasible in theory. We must notice that in a real cell even transitory depletion of the ribosomal pool might be incompatible with cell viability for multiple other reasons not directly related to the bistabiity and collapse of the protein synthesis machinery. Also, the parameter estimations of the protein synthesis model used in this study might be grossly inaccurate. Nevertheless, the theoretical conclusion on the existence of the critical minimum number of ribosomes is independent on the parameter values, and can be potentially validated in an experiment.

The model of ribosomal protein synthesis suggested in this paper needs to be completed with equations describing the synthesis of non-ribosomal proteins. Also, distinguishing 40S and 60S ribosomal proteins might lead to the new interesting dynamical effects such as existence of oscillations in protein synthesis machinery, which potentially can lead to periodic change in the cellular dry mass, even in the absence of proliferation. Exploration of such model extensions, with construction of their complete parametric portrait is beyond the score of this paper but is a feasible though a difficult task.

The main use of simple and basic models of protein synthesis is identification of sensitive (e.g., rate limiting) parameters of the protein synthesis machinery, whose change can efficiently regulate translation. In the past, such an approach was used by us in order to identify the mechanisms of microRNA action by following through the dynamics of the basic observables of translation: amount of mRNA, amount of protein and the polysomal profile, in the presence and in the absence of microRNA. An unsolved inverse problem remains in the case when a regulatory molecule (such as a microRNA) can affect simultaneously several translation parameters. In this case, it is desirable to have an effective mathematical tool allowing "deconvoluting" the mixed effect of the regulation into a vector of strengths of individual translation parameter modulations.

One particular application of the translation models consists in deciphering the results of application of modern sequencing-based technologies quantifying the global state of the translational machinery, such as Ribo-Seq or TRAP-Seq [33]. The amount of data of this type (translatomic data) rapidly grows, but remains less ready for intuitive interpretation compared to other omics data types and require mathematical modeling, taking into account various aspects of polysome [34].

## Acknowledgements

This study was partially supported by ITMO Cancer "Non-coding RNA in cancerology" program (project "Using an integrative strategy to decipher miRNAs function in Pediatric Brain Tumor") and by the Ministry of Science and Higher Education of the Russian Federation (Project No. 14.Y26.31.0022).